\input{epsf}
\documentstyle[editedvolume,numreferences]{crckapb} 

\begin{opening}
\title{The Synergy between Numerical and\protect\\
{\mbox~~~~~}Perturbative Approaches to Black Holes}

\author{Edward Seidel}

\institute{Max-Planck-Institut f{\"u}r Gravitationsphysik,\\
Schlaatzweg 1, 14473 Potsdam, Germany.\\
\\
University of Illinois, Urbana, IL 61801.}

\end{opening}
\runningtitle{Numerical/Perturbative Approaches to Black Holes}

\begin{document}


\begin{abstract}
I describe approaches to the study of black hole spacetimes via 
numerical relativity.  After a brief review of the basic formalisms 
and techniques used in numerical black hole simulations, I discuss a 
series of calculations from axisymmetry to full 3D that can be seen as 
stepping stones to simulations of the full 3D coalescence of two black 
holes.  In particular, I emphasize the interplay between perturbation 
theory and numerical simulation that build both confidence in present 
results and tools to aid and to interpret results of future 
simulations of black hole coalescence.
\end{abstract}
\section{Introduction}
When I started graduate school, I began a thesis project in 
perturbation theory of spherical spacetimes.  I still remember well 
how my advisor, Vincent Moncrief, an expert in perturbation theory, 
advised me to study a paper by Vishveshwara on black hole 
perturbations \cite{Vishveshwara70}.  ``That's the best place to find 
the perturbation formalism'', he told me.  At that time, of course, I 
had no idea how important this subject would continue to be years 
later.  I was very lucky to become grounded in this subject at an 
``early age'', and I knew it would provide important insight into 
problems that were intractable in numerical relativity at that time.  
However I did not appreciate that even as numerical relativity would 
become more and more mature, harnessing hundreds of processors in 
parallel to solve ever larger problems, perturbation theory would 
continue to play such an important role in so many ways.  In fact, its 
role in numerical relativity has become even more important in recent 
years, as I describe below.  Vishu's work in this area influenced me 
in ways that I appreciate even more as my own research moves into 
large scale numerical simulation.

\section{Numerical Evolutions of Black Holes}
The numerical evolution of black holes is very difficult, as one must 
simultaneously deal with singularities inside them, follow the highly 
non-linear regime near the horizons, and also calculate the linear 
regime in the radiation zone where the waves represent a very small 
perturbation on the background spacetime metric.  In axisymmetry this 
has been achieved, for example, for stellar collapse~\cite{Stark85}, rotating 
collisionless matter~\cite{Abrahams94a}, distorted vacuum black holes 
with rotation \cite{Brandt94c} and without \cite{Abrahams92a}, and for 
equal mass colliding black holes \cite{Anninos93b,Anninos94b}, but 
with difficulty.  These 2D evolutions can be carried out to roughly 
$t=100M$, where $M$ is the ADM mass of the spacetime, but beyond 
this point large gradients related to singularity avoiding slicings 
usually cause the codes to become very inaccurate and crash.  This is 
one of the fundamental problems associated with black hole 
evolutions:  if one uses the gauge freedom in the Einstein equations 
to bend time slices up and around the singularities, one ends up with 
pathological behavior in metric functions describing the warped 
slices that eventually leads to numerical instabilities.  

In 3D the problems are even more severe with this traditional, 
singularity avoiding time slicing approach.  To simulate the 
coalescence of two black holes in 3D, evolutions of time scales $t 
\approx 10^{2} - 10^{3}M$ will be required.  Traditional approaches 
can presently carry evolutions only to about $t=50M$.  However, in 
spite of these difficulties, great progress is being made on several 
fronts.  Alternative approaches to standard numerical evolution of 
black holes, such as apparent horizon boundary conditions and 
characteristic evolution, promise much longer evolutions.  Apparent 
horizon conditions cut away the causally disconnected region interior 
to the black hole horizon, allowing better behaved slicings.  These 
have been well developed in 1D, spherically symmetric 
studies~\cite{Seidel92a,Anninos94e,Anninos94c,Scheel94,Marsa96,Cook97a} 
and full 3D evolutions.  Characteristic evolution with ingoing null 
slices have very recently been successful in evolving 3D single black 
holes for essentially unlimited times, even with distortions away from 
spherical or axisymmetry~\cite{Gomez97a}.  These alternate approaches 
look promising, but will take time to develop into general approaches 
to the two black hole coalescence problem, and I will not have space 
to cover them here.  Instead, I focus on how traditional approaches, 
aided by perturbative studies, are providing insight into the dynamics 
of distorted and colliding black holes.  When the alternate techniques 
for black hole evolutions mature, they too will be aided by 
perturbative studies to both verify and interpret numerical 
simulations.

\section{Interplay of Perturbative and Numerical Black Hole Studies}

While numerical relativity is making steady progress in its ability to 
simulate the dynamics of black holes, an equally important and 
exciting development of the last few years has been the application of 
perturbation theory to aid in the verification and interpretation 
numerical simulations.  As I discuss below, there are various ways 
that perturbation theory is used in conjunction with numerical 
relativity, from the extraction of waveforms from fully non-linear 
numerical simulations, to full evolutions of black hole initial data 
sets, treated as perturbations about the Schwarzschild black hole 
background.

In this section I present a series of ideas and calculations that lead 
to what I call a ``Ladder of Credibility''.  Problems need to be 
studied in sequence from easier to more complicated, leading 
ultimately to the 3D spiraling coalescence problem.  I will discuss 
the foundations of black hole perturbation theory, and show how it 
can be used both to test and to interpret results of full numerical 
simulations, beginning with axisymmetric distorted black holes, moving 
on to axisymmetric black hole collisions, and building to full 3D 
simulations.

\subsection{Perturbation Theory}

In this section I give a very brief overview of the theory of 
perturbations of the Schwarzschild spacetime.  This is a topic which 
is very rich and has a long history, and to which Vishu has 
contributed immensely.  A more detailed discussion can be found 
in~\cite{Chandrasekhar83}.  In the next section, I will discuss how we 
apply the theory to study black hole data sets.

One begins the analysis by writing the full metric as a sum of the
Schwarzschild metric $\stackrel{o}{g}_{\alpha\beta}$ and a small perturbation
$h_{\alpha\beta}$:
\begin{equation}
g_{\alpha\beta} = \stackrel{o}{g}_{\alpha\beta} + 
h_{\alpha\beta}.
\end{equation}
One then plugs this expression into the vacuum Einstein equations to 
get equations for the perturbation tensor $h_{\alpha\beta}$.  One 
can separate off the angular part of the solution by expanding 
$h_{\alpha\beta}$ in spherical tensor harmonics, as was originally 
done by Regge and Wheeler~\cite{Regge57}.  For each $\ell,m$ mode, one 
gets separate equations for the perturbed metric functions, which we 
now denote by $h_{\alpha\beta}^{(\ell m)}$.

There are two independent expansions: one, known as even parity, which 
does not introduce any rotational motion in the hole, and one, known as 
odd parity, which does.  We will concentrate here on even parity 
perturbations, as these are the ones which will be relevant for 
studying the data-sets discussed here.  The odd parity perturbations 
produce equations which are very similar, and both can be considered 
in the general case.

We also note here that this treatment is presently restricted to 
perturbations of Schwarzschild black holes.  For the more general 
rotating case, one would like to use the Teukolsky formalism 
describing perturbations of Kerr.  This is much more complicated, and 
has not yet been applied to numerical black hole simulations of the 
kind discussed in this paper.  This is an important research topic 
that needs attention soon!

I have discussed here only the linearized theory of black hole 
perturbations, but this can also be extended to higher order.  This has 
recently been accomplished by Gleiser, Nicasio, Price, and 
Pullin~\cite{Gleiser96a}, who 
worked out equations describing perturbations of Schwarzschild to 
second order in an expansion parameter.  Hence the metric is written 
as 
\begin{equation}
g_{\alpha\beta} = \stackrel{o}{g}_{\alpha\beta} + 
\epsilon h_{\alpha\beta}^{(1)} + \epsilon^{2} h_{\alpha\beta}^{(2)}.
\end{equation}
and the Einstein equations are expanded to second order in $\epsilon$.  
As we will see below, this second order formalism is very useful in 
black hole studies \cite{bht1}.

When dealing with perturbations in relativity, one must be careful 
about interpreting the various metric components $h_{\alpha\beta}$ 
in terms of physics.  Under a coordinate transformation of the form
$x^{\mu}
\rightarrow x^{\mu} + \delta x^{\mu}$, the metric coefficients will 
transform as well.  One can use this gauge freedom to eliminate 
certain metric functions to simplify the corresponding equations for 
the perturbations.  Another, more powerful approach, developed first 
by Moncrief, is to consider linear combinations of the 
$h_{\alpha\beta}$ and their derivatives that are actually {\em 
invariant} under the gauge transformation above.  In either case, the 
analysis leads to a single wave equation for the perturbations of the 
black hole:
\begin{equation}
\label{eq:zerilli}
\frac{\partial^2\psi^{(\ell m)}} {\partial r^{*2}} -
\frac{\partial^2\psi^{(\ell m)}}{\partial t^2} +
V^{(\ell)}\left(r\right) \psi^{(\ell m)} = 0,
\end{equation}
where the potential function $V^{(\ell)}(r)$ is given by
\begin{eqnarray}
V^{(\ell)}\left(r\right) = \left( 1 - \frac{2M}{r} \right) \times
\hspace{3.75in}
\nonumber \\
\left\{
  \frac{1}{\Lambda^2} \left[ \frac{72M^3}{r^5} - \frac{12M}{r^3}
    \left( \ell-1 \right) \left( \ell+2\right) \left( 1-\frac{3M}{r}
    \right) \right] +
 \frac{\ell
    \left(\ell-1\right)\left(\ell+2\right)
    \left(\ell+1\right)}{r^2\Lambda} \right\},
\end{eqnarray}
where $r$ is the standard Schwarzschild radial coordinate, and $r^*$ 
is the so-called tortoise coordinate, given by $r^* = r + 
2M\ln\left(r/2M-1\right)$, and $\Lambda$ is a function of $r$ 
described below.  For all $\ell$, the potential function is 
positive and has a peak near $r=3M$.  This equation was first derived 
by Zerilli \cite{Zerilli70}.  Regge and Wheeler found an analogous 
equation for the odd parity perturbations, which are much simpler than 
the even parity perturbations, 13 years earlier in 
1957~\cite{Regge57}.  Note that the potential depends on $\ell$, but 
is independent of $m$.  This remarkable equation (along with its 
odd-parity counterpart, known as the Regge-Wheeler equation) completely 
describes gravitational perturbations of a Schwarzschild black hole.  
It continues to be studied by many researchers now 40 years after this 
perturbation program was begun in 1957.  

When carried out to second order in the perturbation parameter, the 
first order equations are of course the same, but they are accompanied 
by an equation representing the second order correction.  Remarkably, 
the second order equation has {\em exactly the same form} as the first 
order equation, with source terms which are non-linear in the first 
order perturbation quantities.  As I mention below, the second order 
perturbation theory has been used in application to the head-on 
collision of two black holes to extend the range of validity of the 
first order treatment.  But perhaps its most important use is in 
providing a check on the validity of the first order perturbation 
treatment.  When second order corrections are small compared to first 
order results, one expects the first order results to be reliable.

Using the first order equation, it has been shown that Schwarzschild 
is stable to perturbations, and has characteristic oscillation 
frequencies known as quasinormal modes \cite{bht2}.  These modes are solutions to 
the Zerilli equation as given in Eq.~(\ref{eq:zerilli}) which are 
completely ingoing at the horizon ($r^*=-\infty$) and completely 
outgoing at infinity ($r^*=\infty$).  For each $\ell$-mode, 
independent of $m$, there is a fundamental frequency, and overtones.  
These are very important results!  One expects that a black hole, when 
perturbed in an arbitrary way, will oscillate at these quasinormal 
frequencies.  This will give definite signals to look for with 
gravitational wave observatories.

These quasinormal frequencies are complex, meaning they have an 
oscillatory and a damping part (not growing---black holes are 
stable!), so the oscillations die away as the waves carry energy away
from the system.  The frequencies depend only on the mass, spin, and 
charge of the black hole.

There are numerous ways in which this perturbation theory has become 
essential in numerical black hole simulations, and the rest of this 
paper will concentrate on this subject.  First of all, the fact that 
perturbation theory reveals that black holes have quasinormal mode 
oscillations raises expectations about the evolution of distorted 
black holes: they should, at least in the linear regime, oscillate at 
these frequencies which should be seen in fully non-linear numerical 
simulations.  But are they still seen in highly non-linear interactions, 
e.g., in the collision of two black holes?  Secondly, as we will see, 
this perturbation theory provides a method by which to separate out the 
Schwarzschild background from the wave degrees of freedom, which can 
be used to find waves in numerical simulations.  Finally, as the 
perturbations are governed by their own evolution equation, this 
equation should be useful to actually evolve some classes of black 
hole initial data, as long as they represent slightly perturbed black 
holes, and this can be used as an important check of fully non-linear 
numerical codes.  Certainly, during the late stages of black hole 
coalescence, the system will settle down to a slightly perturbed 
black hole, and numerical codes had better be able to accurately 
compute waves from such systems if they are to be used to help 
researchers find signals in actual data collected by gravitational 
wave observatories.

\subsection{Waveform Extraction}

In this section I show how to take this perturbation theory and apply 
it in a practical way to numerical black hole simulations.  One 
considers now the numerically generated metric $g_{\alpha\beta,{\rm num}}$ 
to be the sum of a spherically symmetric part and a perturbation: 
$g_{\alpha\beta,{\rm num}}=\stackrel{o}{g}_{\alpha\beta}+h_{\alpha \beta}$, 
where the perturbation $h_{\alpha\beta}$ is expanded in tensor 
spherical harmonics as before.  To compute the elements of 
$h_{\alpha\beta}$ in a numerical simulation, one integrates the 
numerically evolved metric components $g_{\alpha\beta,{\rm num}}$ against 
appropriate spherical harmonics over a coordinate 2--sphere 
surrounding the black hole.  The orthogonality of the $Y_{\ell m}$'s 
allows one to ``project'' the contributions of the general wave signal 
into individual modes, as explained below.  The resulting functions 
can then be combined in a gauge-invariant way, following the 
prescription given by Moncrief \cite{Moncrief74}, leading directly to 
the Zerilli function.  This procedure was originally developed by 
Abrahams \cite{Abrahams88} and developed further by various groups.

As mentioned above, we assume the general metric can be decomposed into its
spherical and non-spherical parts.  We assume here that the 
background is Schwarzschild, written in Schwarzschild coordinates, but 
the treatment of a more general spherical background in an arbitrary 
coordinate system is possible~\cite{Seidel90c}.
The nonspherical perturbation tensor $h_{\mu\nu}$ 
for even-parity perturbations can be written
\begin{eqnarray}
h_{tt} &=& -S H_0^{(\ell m)} Y_{\ell m} \\
h_{tr} &=& H_1^{(\ell m)} Y_{\ell m} \\
h_{t\theta} &=& h_0^{(\ell m)} Y_{\ell m,\theta} \\
h_{t\phi} &=& h_0^{(\ell m)} Y_{\ell m,\phi} \\
h_{rr} &=& S^{-1} H_2^{(\ell m)} Y_{\ell m} \\
h_{r\theta} &=& h_1^{(\ell m)} Y_{\ell m,\theta} \\
h_{r\phi} &=& h_1^{(\ell m)} Y_{\ell m,\phi} \\
h_{\theta\theta} &=& r^2 K^{(\ell m)} Y_{\ell m} + r^2 G^{(\ell m)} Y_{\ell m,\theta\theta}
\\
h_{\theta\phi} &=& r^2 G^{(\ell m)} \left( Y_{\ell m,\theta\phi} - \cot\theta Y_{\ell
    m,\phi} \right) \\
h_{\phi\phi} &=& r^2 K^{(\ell m)} \sin^2\theta Y_{\ell m} + r^2 G^{(\ell m)} \left( Y_{\ell
    m,\phi\phi} + \sin\theta \cos\theta Y_{\ell m,\theta} \right)
\end{eqnarray}
where $S = (1-2M/r)$ comes from the Schwarzschild background, and $r$ 
is the standard Schwarzschild radial coordinate.  I have taken these 
expressions directly from Vishu's 1970 paper \cite{Vishveshwara70} 
where I first learned them.

Each $\ell m$-mode of $h_{\mu\nu}$ can then be obtained numerically by 
projecting the full metric against the appropriate $Y_{\ell m}$.  So, 
for example,
\begin{equation}
H_2^{(\ell m)} = (1-2M/r) \int g_{rr, {\rm num}} Y_{\ell m} d\Omega .
\end{equation}
More complex expressions can be developed for all metric functions, and 
on a more general background, as detailed 
in~\cite{Allen97a,Allen98a}.  These functions can be computed 
numerically, given a numerically generated spacetime, as described 
below.

Once the perturbation functions $H_{2}, K, G, etc.$ are obtained, one 
still needs to create the special combination that obeys the Zerilli 
equation.  Moncrief showed that the Zerilli function that obeys this 
wave equation is gauge invariant in the sense discussed above, and can 
be constructed from the Regge-Wheeler variables as follows:
\begin{equation}
\psi^{(\ell m)} = \sqrt{\frac{2(\ell-1)(\ell+2)}{\ell (\ell+1)}} \frac{4 r S^2
k^{(\ell m)}_2 + \ell (\ell+1) r k^{(\ell m)}_1}{\Lambda},
\end{equation}
where
\begin{eqnarray}
\Lambda &\equiv& \ell (\ell+1) - 2 + \frac{6M}{r} \\
k^{(\ell m)}_1 &\equiv& K^{(\ell m)} + S r G^{(\ell m)}_{,r} - 2 \frac{S}{r}
h^{(\ell m)}_1 \\
k^{(\ell m)}_2 &\equiv& \frac{H_2^{(\ell m)}}{2S} - \frac{1}{2\sqrt{S}} \frac{\partial}
{\partial r} \left( \frac{rK^{(\ell m)}}{\sqrt{S}} \right) \\
S &\equiv& 1 - \frac{2M}{r}.
\end{eqnarray}

In order to compute the Regge-Wheeler perturbation functions $h_1$, 
$H_2$, $G$, and $K$, one needs the spherical metric functions on some 
2-sphere.  We get these in 3D by interpolating the Cartesian metric 
functions onto a surface of constant coordinate radius, and computing 
the spherical metric functions from these using the standard 
transformation. These are straightforward but messy calculations 
which are covered in detail in Refs. \cite{Allen97a,Allen98a}.

In summary, in this section I have outlined a practical approach to 
the use of perturbation theory as a tool to construct a gauge 
invariant measure of the gravitational radiation in a numerically 
generated black hole spacetime. There are various ways in which this 
information can be used, which is the subject of the following sections.

\subsection{Applications}

In the sections above, I showed how to extract the gauge invariant 
Zerilli function at a given radius on some time slice of the numerical 
spacetime.  In the following sections I show several ways in which 
this information can be used, including (a) {\em evolving} a 
numerically (or analytically) generated initial data set with 
perturbation theory and (b) extracting waveforms from a fully 
non-linear evolution, possibly from the same data set.  This 
perturbative approach has also recently been successful in providing
outer boundary conditions for a 3D numerical 
simulation \cite{Abrahams97a}.

\subsubsection{Axisymmetric Distorted Black Holes}

We begin with the study of axisymmetric single black holes that have 
been distorted by the presence of an adjustable torus of non-linear 
gravitational waves which surround it.  The amplitude and shape of the 
initial wave can be specified by hand, as described below, and can create 
very highly distorted black holes.  Such initial data sets, and their 
evolutions in axisymmetry, have been studied extensively, as described 
in Refs. \cite{Abrahams92a,Bernstein94a,Bernstein93b}.  For our 
purposes, we consider them as convenient initial data that create a 
distorted black hole that mimics the merger, just after coalescence, 
of two black holes colliding in axisymmetry \cite{Anninos94b}.

Following \cite{Bernstein94a}, we write the 3--metric
in the form originally used by Brill~\cite{Brill59}:
\begin{equation}
\label{eq:metric}
d\ell^2 = \tilde{\psi}^4 \left( e^{2q} \left( d\eta^2 + d\theta^2 \right) +
  \sin^2\theta d\phi^2 \right),
\end{equation}
where $\eta$ is a radial coordinate related to the standard Schwarzschild 
isotropic radius $\bar{r}$ by $\bar{r} = e^{\eta}$. (We have set
the scale parameter $m$ in \cite{Bernstein94a} to be 2 in this paper.) 
We choose our initial slice to be time symmetric, so that the
extrinsic curvature vanishes. Thus, given a
choice for the ``Brill wave'' function $q$, the Hamiltonian constraint
leads to an elliptic equation for the conformal factor $\tilde{\psi}$.  The
function $q$ represents the gravitational wave surrounding the black
hole, and is chosen to be
\begin{equation}
\label{eq:q2d}
q\left(\eta,\theta,\phi\right) = a \sin^n\theta \left(
  e^{-\left(\frac{\eta+b}{w}\right)^2} +
  e^{-\left(\frac{\eta-b}{w}\right)^2} \right)
  \left(1+c \cos^2\phi\right).
\end{equation}
Thus, an initial data set is characterized by the parameters 
$\left(a,b,w,n,c\right)$, where, roughly speaking, $a$ is the 
amplitude of the Brill wave, $b$ is its radial location, $w$ its 
width, and $n$ and $c$ control its angular structure.  A study 
of full 3D initial data and their evolutions are discussed elsewhere 
 \cite{Camarda97a,Camarda97c,Brandt97a}.  If the amplitude $a$ 
vanishes, the undistorted Schwarzschild solution results, leading to
\begin{equation}
\tilde{\psi} = 2 \cosh \left( \frac{\eta}{2} \right),
\end{equation}
which puts the metric (\ref{eq:metric})in the standard Schwarzschild isotropic form.

\paragraph{Linear Evolution}

We now consider the evolution of these distorted black holes.  If the 
amplitude is low enough, this should represent a small perturbation on 
a Schwarzschild black hole, and hence the system should be amenable 
to a perturbative treatment.  The idea is to actually compute 
$\psi(r,t=0)$, by extracting the Zerilli function at every radial grid 
point on the initial data slice, and use the Zerilli evolution 
equation to actually evolve the system as a perturbation.  This will 
allow us to compare the waveform at some radius $\psi_{lin}(r_{0},t)$ 
obtained in this way, to that obtained with a well-tested 2D 
axisymmetric code that performs full non-linear evolutions.  This will 
serve as a test of the initial Zerilli function being given to the 
linear code, and of the linear evolution code itself.  It will also 
help us determine for which Brill wave amplitudes this procedure 
breaks down.  Beyond a certain point, perturbation theory will fail 
and non-linear effects will become important.

\begin{figure}
\epsfxsize=200pt \epsffile{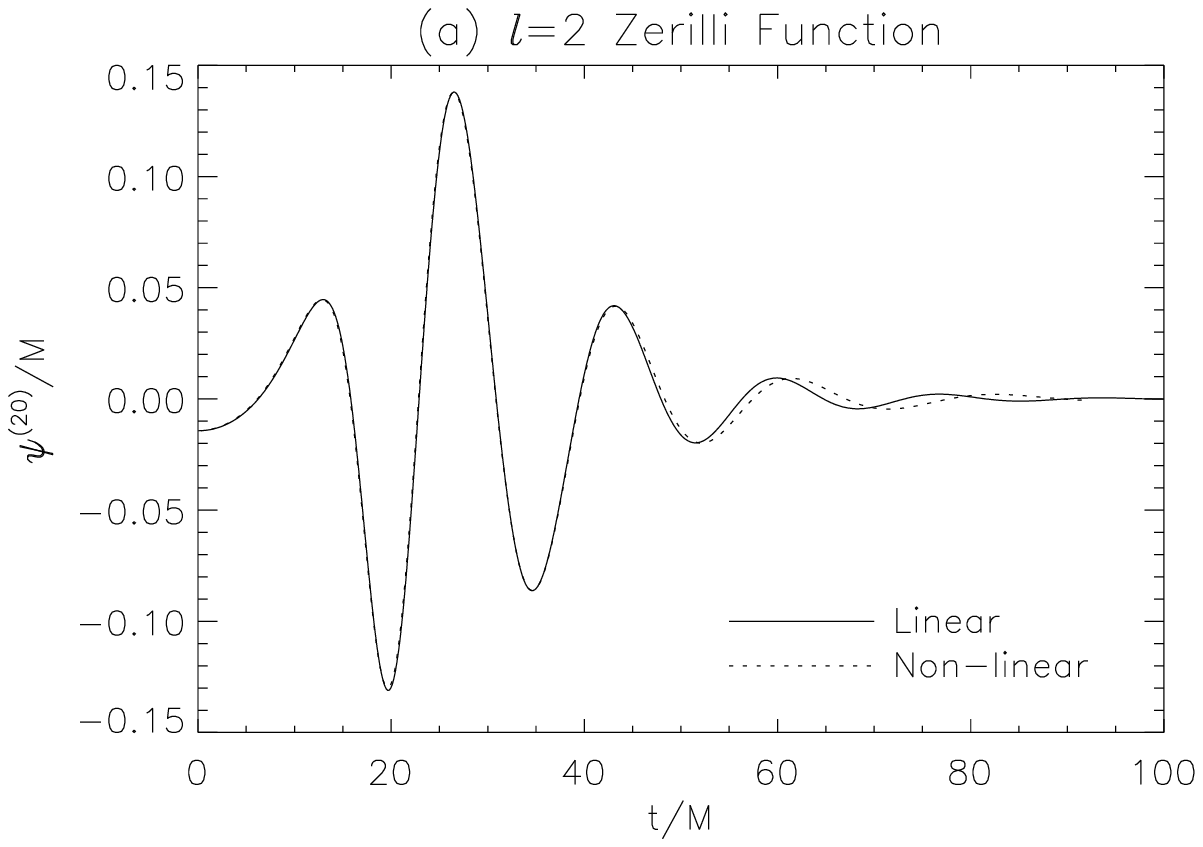}
\epsfxsize=200pt \epsffile{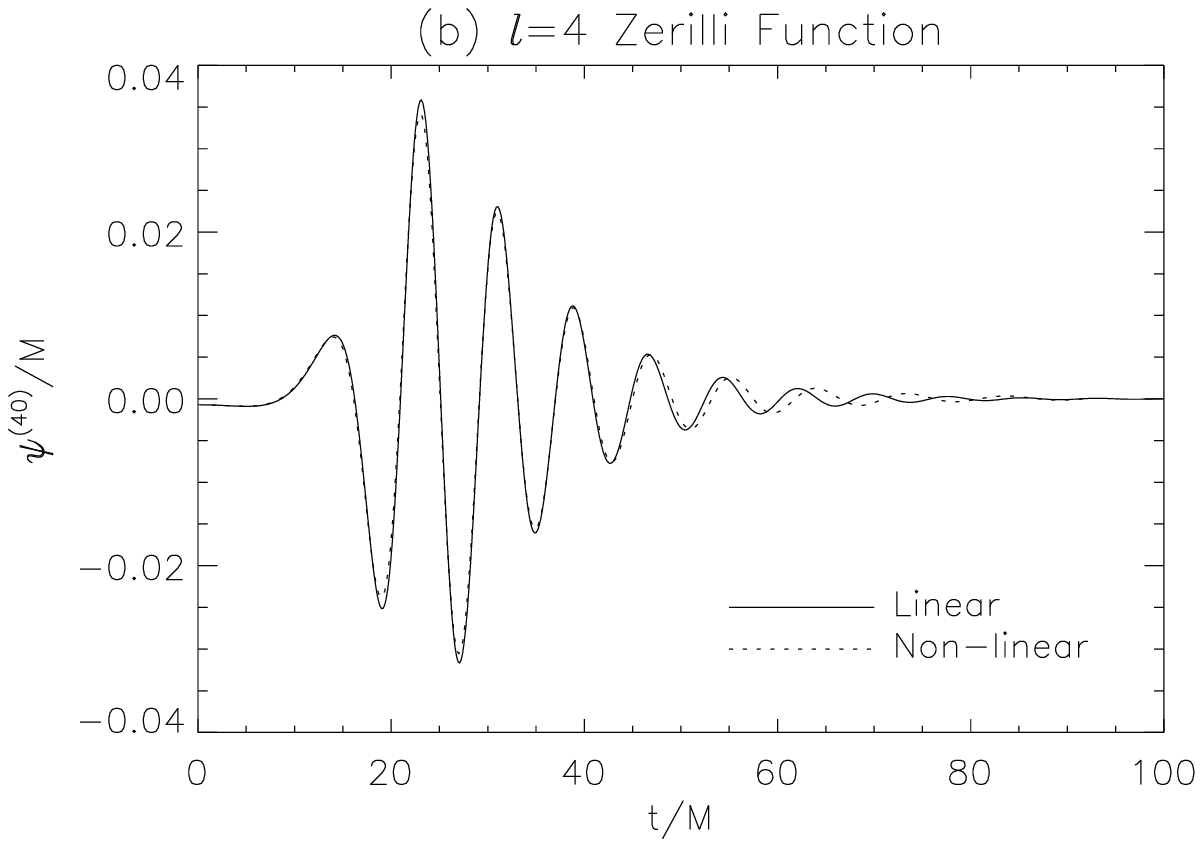}
\caption{We show the (a)$\ell=2$ and (b)$\ell=4$ Zerilli functions as a
function of time, extracted during linear and 2D non-linear evolutions of the
data set $(a,b,w,n,c)=(0.05,1,1,4,0)$. The data were extracted at a radius of
$r=15M$.}
\label{fig:run1}
\end{figure}

Let us first consider the data set $(a,b,w,n,c)=(0.05,1,1,4,0)$, in the 
notation above.  In this case the Brill wave is 
initially far from the black hole, and will propagate in, hitting it 
and exciting the normal mode oscillations.  In Figure~\ref{fig:run1} 
we show the $\ell=2$ and $\ell=4$ Zerilli functions as a function of 
time, at a radius of $r=15M$.  Data are shown from both the linear and 
2D non-linear codes.  We see that for both functions, the linear and 
non-linear results line up nicely until about $t=50M$, when a phase 
shift starts to be noticeable.  This phase shift and widening of the 
wave at late times is known from previous studies of numerical 
simulations of distorted black hole spacetimes in 
axisymmetry~\cite{Abrahams92a}.

\begin{figure}
\epsfxsize=200pt \epsffile{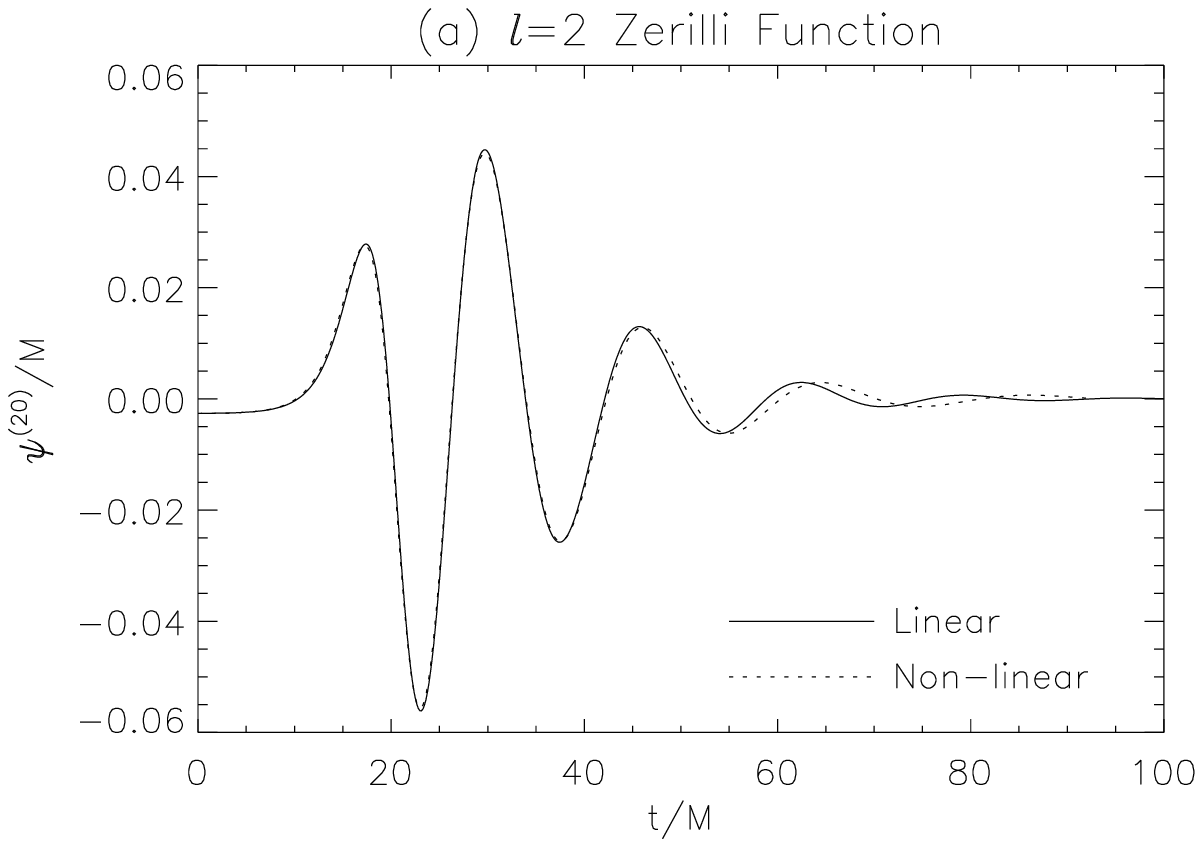}
\epsfxsize=200pt \epsffile{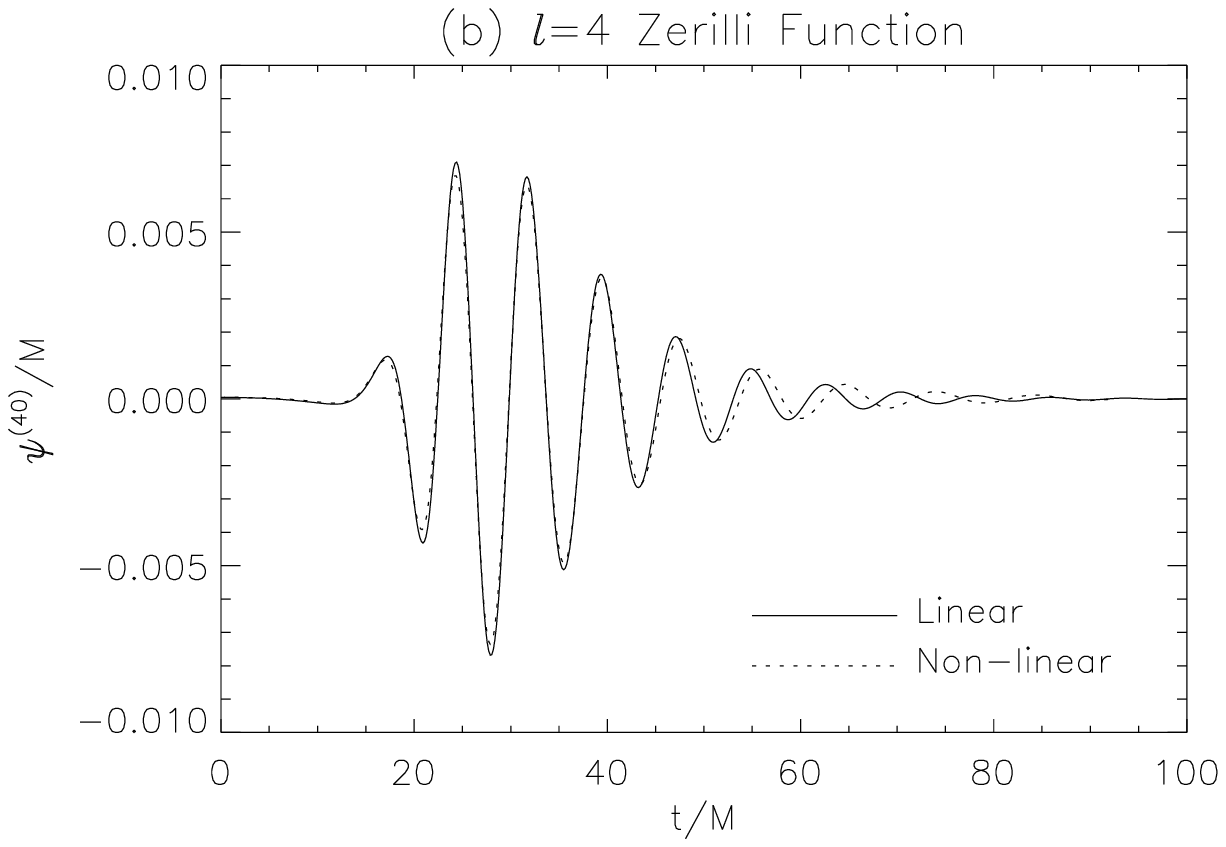}
\caption{We show the (a)$\ell=2$ and (b)$\ell=4$ Zerilli functions as a
function of time, extracted during linear and 2D non-linear evolutions of the
data set $(a,b,w,n,c)=(0.05,0,1,4,0)$. The data were extracted at a radius of
$r=15M$.}
\label{fig:run2}
\end{figure}

As second example, let us look at $(a,b,w,n,c)=(0.05,0,1,4,0)$.
In this case the Brill wave is initially right on the throat.
In Figure~\ref{fig:run2} we show the $\ell=2$ and the $\ell=4$
waveforms as a function of time extracted at a radius of $r=15M$.
Again, data from both the linear and 2D non-linear codes are shown. The
data line up well until about $t=50M$, when phase errors again show up.

This is the first step in our ladder of credibility.  We have shown 
that these data sets provide an important testbed for a numerical 
black hole evolution.  First, they confirm that these data sets can be 
treated as linear perturbations on a Schwarzschild background, since 
both linear and fully non-linear evolutions agree.  Second, they 
remarkably confirm the results of a complex, non-linear evolution code 
which evolves a black hole (in maximal slicing).  This gives great 
confidence in the ability of this code to treat black holes and 
extract waveforms, even in the more highly distorted cases where 
perturbation theory breaks down (but waveform extraction will not 
necessarily break down, at least far from the hole).  We will use this 
technique in various ways below.

\subsubsection{Axisymmetric Black Hole Collisions}

We now turn to another application of this basic idea of evolving 
dynamic black hole spacetimes with perturbation theory, but this time 
we consider two black holes colliding head on.  It might seem to be 
impossible to treat colliding black holes perturbatively, but there 
are two limits in which perturbation theory has been shown to be 
incredibly successful.  First, if the two holes are so close together 
initially that they have actually already merged into one, they might 
be considered as a single perturbed Schwarzschild hole (the so-called 
``close limit'').  Using similar ideas to those discussed above, Price 
and Pullin and others 
 \cite{Price94a,Price94b,Abrahams95c,Baker96a,Gleiser96b} used this 
technique to produce waveforms for colliding black holes in the 
Misner \cite{Misner60} and Brill and Lindquist \cite{Brill63} black hole 
initial data.  These initial data sets for multiple black holes are 
actually known analytically.  The original paper of Price and Pullin 
 \cite{Price94a} is what spurred on so much interest in these many 
applications of perturbation theory as a check on numerical 
relativity.  Second, when the holes are very far apart, one can 
consider one black hole as a test particle falling into the other.  
Then one rescales the answer obtained by formally allowing the ``test 
particle" to be a black hole with the same mass as the one it is 
falling into \cite{Anninos93b,Anninos94b,Price94b}.

The details of this success has provided insights into the nature of 
collisions of holes, and should also apply to many systems of 
dynamical black holes.  The waveforms and energies agree remarkably 
well with numerical simulations.  Moreover, second order perturbation 
theory \cite{Gleiser96b} spectacularly improved the agreement between 
the close limit and full numerical results for even larger 
distances between the holes, although ultimately beyond a certain 
limit the approximation is simply inappropriate and breaks down.

The success of these techniques suggests, among other things, that 
these are very powerful methods that can be used hand-in-hand with 
fully non-linear numerical evolutions, and can be applied in a variety 
of black hole spacetimes where one might naively think they would not 
work.  For these reasons, many researchers are continuing to apply 
these techniques to the axisymmetric case with more and more 
complicated black hole spacetimes (e.g., the collision of boosted 
black holes \cite{Baker96a}, or
counter-rotating, spinning black holes, colloquially known as the 
``cosmic screw''.)  Furthermore, these techniques will become even 
more essential in 3D, where we cannot achieve resolution as high as we 
can with 2D codes.

Finally, this is yet another rung on the ``ladder of credibility'':  
we now have not only slightly perturbed Schwarzschild spacetimes to 
consider, but also a series of highly nontrivial colliding black hole 
spacetimes that are now well understood in axisymmetry due to the nice 
interplay between perturbative and fully numerical treatments of the 
same problems.  These then provide excellent testbeds for 3D 
simulations, which we turn to next.

\subsubsection{3D Testbeds}

Armed with robust and well understood axisymmetric black hole codes, 
we now consider the 3D evolution of axisymmetric distorted black hole 
initial data.  These same axisymmetric initial data sets can be ported 
into a 3D code in cartesian coordinates, evolved in 3D, and the 
results can be compared to those obtained with the 2D, axisymmetric 
code discussed above.  The 3D code used to evolve these black hole 
data sets is described in Refs. \cite{Anninos94c,Camarda97b}, and the 
simulations described here are major simulations on very large 
supercomputers:  they require about 12 GBytes of memory and take more 
than 24 hours on a 128 processor SGI Origin 2000 computer.

\begin{figure}
\epsfxsize=200pt \epsffile{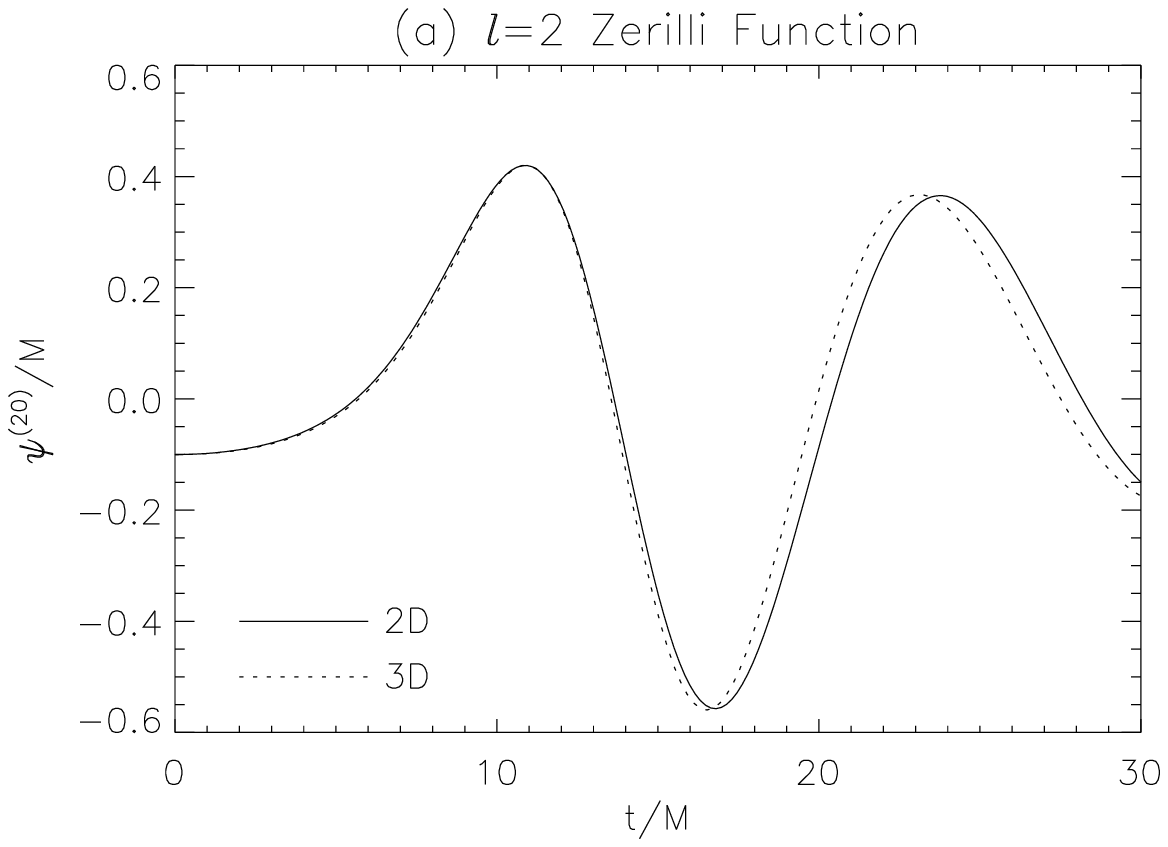}
\epsfxsize=200pt \epsffile{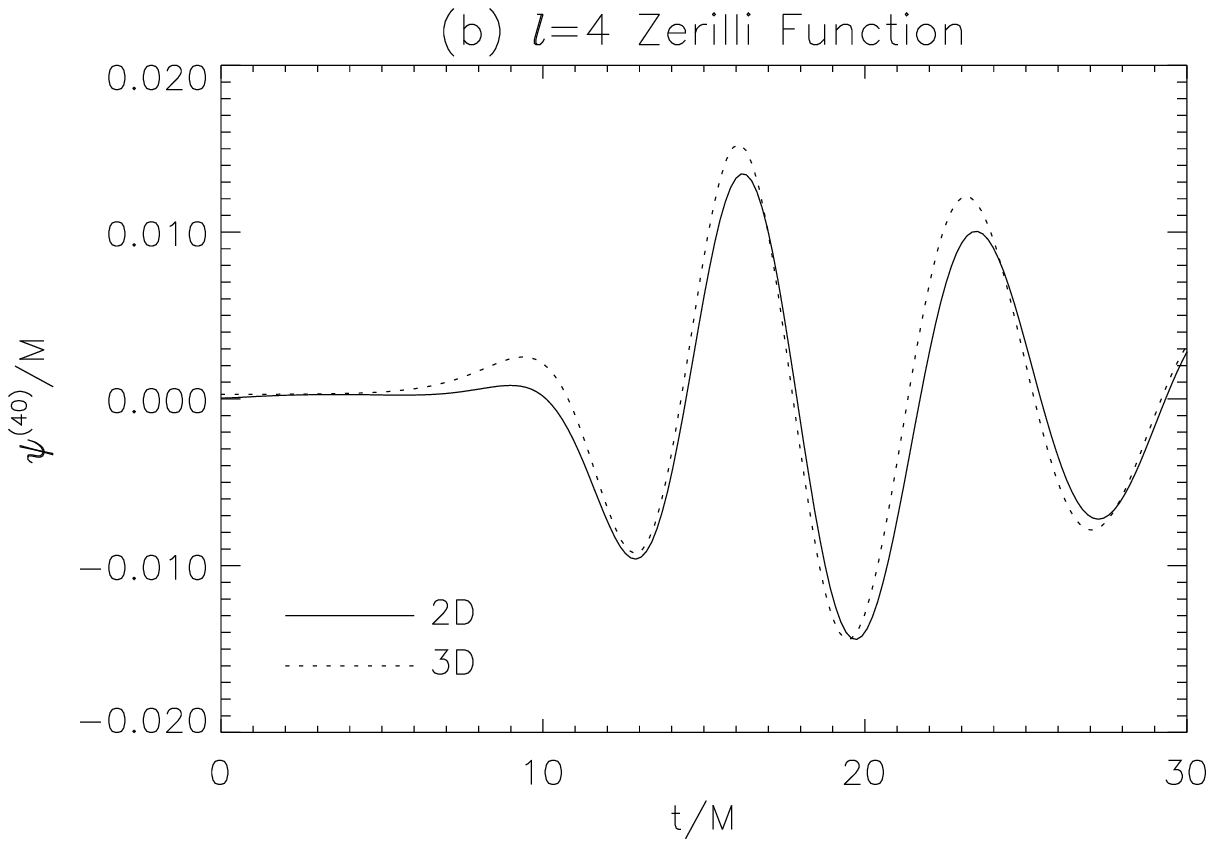}
\caption{We show the (a) $\ell=2$ and (b) $\ell=4$ Zerilli functions
vs. time, extracted during 2D and 3D evolutions of the data set
$(a,b,w,n,c)=(0.5,0,1,2,0)$. The functions were extracted at a radius of
$8.7M$. The 2D data were obtained with $202\times 54$ grid points,
giving a resolution of $\Delta\eta=\Delta\theta=0.03$. The 3D data
were obtained using $300^3$ grid points and a resolution of $\Delta
x=0.0816M$.}
\label{fig:zer2dcomp}
\end{figure}

In the first of these simulations, we study the evolution of the 
distorted single black hole initial data set $(a,b,w,n,c) = 
(0.5,0,1,2,0)$.  As the azimuthal parameter $c$ is zero, this is 
axisymmetric and can also be evolved in 2D. In 
Figure~\ref{fig:zer2dcomp}a we show the result of the 3D evolution, 
focusing on the $\ell=2$ Zerilli function extracted at a radius 
$r=8.7M$ as a function of time.  Superimposed on this plot is the same 
function computed during the evolution of the same initial data set 
with a 2D code, based on the one described in detail in 
 \cite{Abrahams92a,Bernstein93b}.  The agreement of the two plots over 
the first peak is a strong affirmation of the 3D evolution code and 
extraction routine.  It is important to note that the 2D results were 
computed with a different slicing (maximal), different coordinate 
system, and a {\em different spatial gauge}.  Yet the physical results 
obtained by these two different numerical codes, as measured by the 
waveforms, are remarkably similar (as one would hope).  A full 
evolution with the 2D code to $t=100M$, by which time the hole has 
settled down to Schwarzschild, shows that the energy emitted in this 
mode at that time is about $4\times 10^{-3}M$.  This result shows that 
now it is possible in full 3D numerical relativity, in cartesian 
coordinates, to study the evolution and waveforms emitted from highly 
distorted black holes, even when the final waves leaving the system 
carry a small amount of energy.

In Fig.~\ref{fig:zer2dcomp}b we show the $\ell=4$ Zerilli function 
extracted at the same radius, computed during evolutions with 2D and 
3D codes.  This waveform is more difficult to extract, because it has 
a higher frequency in both its angular and radial dependence, and it 
has a much lower amplitude: the energy emitted in this mode is about three 
orders of magnitude smaller than the energy emitted in the $\ell=2$ 
mode, yet it can still be accurately evolved 
and extracted.  This is quite a remarkable result, and bodes well for 
the ability of numerical relativity codes ultimately to compute 
accurate waveforms that will be of great use in interpreting data 
collected by gravitational wave detectors.  (However, as I point out 
below, there is a quite a long way to go before the general 3D 
coalescence can be studied!)

These results have been reported in much more detail in 
 \cite{Camarda97a,Camarda97b}.

\subsubsection{True 3D Distorted Black Holes}

We now turn to radiation extraction in true 3D black hole evolutions.  
This is of major importance for the connection between numerical 
relativity and gravitational wave astronomy.  Gravitational wave 
detectors such as LIGO, VIRGO, and GEO will measure these waves 
directly, and may depend on numerical relativity to provide templates 
to both extract the signals from the experimental data and to 
interpret the results.  

In the sections above, I showed by comparison to 2D results that a 3D 
code is able to accurately simulate distorted black holes.  Armed with 
these tests, we now consider evolutions of initial data sets which are 
non-axisymmetric distorted black holes, {\it i.e.}, data sets which 
have non-vanishing azimuthal parameter $c$.  We also consider the 
evolution of the same distorted black hole data sets via linearized 
theory, as we did with the 2D results presented above.  The techniques 
are the same, although the details in a full 3D treatment are more 
complicated.  Please refer to Refs. \cite{Allen97a,Allen98a} for more 
details.

\begin{figure}
\epsfxsize=4in \epsffile{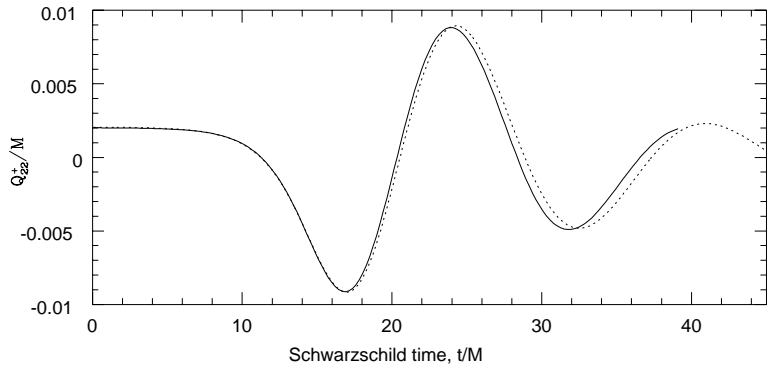}
\caption{\label{fig3d_17}
We show waveforms for the $\ell=m=2$ Zerilli function extracted from the 
linear and non-linear evolution codes for a fully 3D, nonaxisymmetric 
distorted black hole.  The dotted line shows the linear evolution, 
evolving only the Zerilli equation, and the solid line shows the non-linear 
evolution in full 3D cartesian coordinates, using a massively parallel 
supercomputer.}
\end{figure}

The initial data set $(a=-0.1,b=0,c=0.5,w=1,n=4)$ was evolved with the 
3D numerical relativity Cartesian code described above, with $300^{3}$ 
grid zones points in each coordinate direction.  In 
Figure~\ref{fig3d_17} we show the $\ell=m=2$ Zerilli function computed 
during the evolution, comparing both the full non-linear theory with 
the linearized treatment.  This is the first time a 3D non-linear 
numerical relativity code has been used to compute waveforms from fully 
3D distorted black hole (but also see recent results using a 
characteristic formulation~\cite{Gomez97a}). From the figure it is 
clear that the 3D results agree well with the perturbative treatment, 
even though the energy carried by these waves is very small.

Nonaxisymmetric modes are extracted not just because they {\em can} 
be, but because they should be quite {\em important} for gravitational 
wave observatories.  It turns out that the $\ell=2,m=2$ is considered 
to be one of the most promising black hole modes to be seen by 
gravitational wave detector.  In realistic black hole coalescence, the 
final hole is expected to have a large amount of angular momentum, 
possibly near the Kerr limit $a=1$.  This particular mode is one of 
the least damped (much less damped than for Schwarzschild, as seen 
here), and is also expected to be strongly excited \cite{Flanagan97a}.  
Therefore, it is important to begin exhaustively testing the code's 
ability to generate and cleanly extract such nonaxisymmetric modes, 
even in the case studied here without rotation.

Many more modes can be extracted, including $\ell=4,m$ modes, and 
details and analysis can be found in 
 \cite{Camarda97a,Camarda97d,Allen97a,Allen98a}. Comparisons between 
the perturbative and non-linear results reveal that waveforms can be 
accurately extracted in the linear regime, and that even at very low 
distortion amplitudes non-linear effects appear.  The comparisons with
perturbation theory are essential in understanding these effects, and 
will continue to be for some years to come.

\section{Summary}

I have given a brief overview of work on evolutions of distorted 
black holes and black hole collisions over the last decade, from 2D 
axisymmetric studies to recent 3D studies.  At each stage along the 
way, perturbation theory has turned out to be an essential ingredient 
in the program.  Our understanding of the 2D collision of two black 
holes has been aided immensely from perturbation theory, and in the 
last year our ability to simulate true 3D distorted black holes, which model 
the late stages of 3D binary black hole coalescence, has matured 
considerably.  

Unfortunately, we still have a very long way to go!  Although one can 
now do 3D evolutions of distorted black holes, and accurately extract 
very low amplitude waves, the calculations one can 
presently do are actually very limited.  With present techniques, the 
evolutions can only be carried out for a fraction of the time required 
to simulate the 3D orbiting coalescence.  Most of what has been 
described here has been with certain symmetries, or with a single 
black hole in full 3D. At the present time, I am only aware of one 
attempt to study the collision of two black holes in 3D without any 
symmetries, which was recently recently reported by Br{\"u}gmann 
 \cite{Bruegmann97}.  However, this calculation is treated as a 
feasibility study, without detailed waveform extraction at this point, 
and again the evolution times are quite limited (less than those 
reported here.)

Many new techniques are under development to extend the simulations to 
the time scales required for true binary black hole coalescence, such 
as apparent horizon boundary conditions \cite{Seidel92a,Cook97a}, 
hyperbolic systems \cite{Bona97a,Abrahams95a}, and characteristic 
evolution \cite{Gomez97a}.  Many of these (or perhaps all!)  may be 
needed to handle the general, long term evolution of coalescing black 
holes.  Each of these techniques may introduce numerical artifacts, 
even if at very low amplitude, to which the waveforms may be very 
sensitive.  As new methods are developed and applied to numerical 
black hole simulations, they can now be tested on evolutions such as 
those presented here to ensure that the waveforms are accurately 
represented in the data.

In closing, I want to emphasize that the kind of work that Vishu 
helped to pioneer 30 years ago is still at the forefront of numerical 
simulations aided by the world's most powerful computers, and it 
seems clear that perturbation theory will continue to play an 
essential role in both the verification and physical understanding of 
large scale numerical simulations.  There is still much work to do in 
this area, both on the theoretical and numerical areas.  I expect it 
will continue to excite researchers for years to come!

\section{Acknowledgments}

This work has been supported by the Albert Einstein Institute (AEI) 
and NCSA. I am most thankful to Karen Camarda and Gabrielle Allen for 
carrying out much of the 3D work reported here.  
Among many colleagues who have contributed to this work on black 
holes, I also thank Andrew Abrahams, Pete Anninos, 
David Bernstein, Steve Brandt, David Hobill, Joan Mass{\'o}, John 
Shalf, Larry Smarr, Wai-Mo Suen, John Towns, and Paul Walker.

I would like to thank K.V. Rao and the staff at NCSA for 
assistance with the computations.  Calculations were performed at AEI 
and NCSA on an SGI/Cray Origin 2000 supercomputer.

%

\end{document}